# SYSTEM-LEVEL GENETIC CODES: AN EXPLANATION FOR BIOLOGICAL COMPLEXITY

By John F. McGowan, Ph.D.
Desktop Video Expert Center
NASA Ames Research Center
Mail Stop 233-18
Moffett Field, CA 94035-1000
E-Mail: jmcgowan@mail.arc.nasa.gov
Telephone: (650) 604-0143

(2/24/00 2:41 PM)

## ABSTRACT

Complex systems with tightly coadapted parts frequently appear in living systems and are difficult to account for through Darwinian evolution, that is random variation and natural selection, if the constituent parts are independently coded in the genetic code. If the parts are independently coded, multiple simultaneous mutations appear necessary to create or modify these systems. It is generally believed that most proteins are independently coded. The textbook rule is one gene for one enzyme. Thus, biochemical systems with tightly coadapted parts such as the blood clotting cascade pose a difficulty for Darwinian evolution. This problem can be overcome if the current understanding of the genetic code is incomplete and a system-level genetic code in which seemingly independent proteins are encoded in an interdependent, highly correlated manner exists. The methods by which human beings design and fabricate complex systems of tightly coadapted parts are explored for insights into the requirements for a system-level genetic code. Detailed examples of system-level codes for networks of matching parts are presented. The implications of identifying and deciphering the system-level genetic code if it exists for the prevention, treatment, and cure of heart disease, cancer, immune disorders, and for rational drug design are discussed.

1. INTRODUCTION

An irreducibly complex system is a system of tightly coadapted interacting parts where the removal or even slight modification of any single part breaks the system[1]. A lock and key is a simple example of an irreducibly complex system. The lock is useless without the key. The key is useless without the lock. The lock and the key are tightly matched. A small change in the shape of the key or the lock results in the key failing to open the lock. This tight coadaptation of parts is common in man-made machines and appears necessary for many machines to function.

A wide variety of biomechanical and biochemical systems appear to be irreducibly complex. Charles Darwin referred to this problem as organs of extreme complexity in the context of biomechanical systems. The proteins comprising complex biochemical systems are frequently tightly matched. For example, a common constituent of biochemical systems is a matched pair of proteins, an activation enzyme and a proenzyme. The activation enzyme cleaves the otherwise inert proenzyme yielding an active enzyme. The activation enzyme cleaves only the proenzyme. The activation enzyme is useless without the proenzyme. The proenzyme is useless without the activation enzyme. The two proteins are tightly matched so that a small change in either part will break the system. The blood clotting system in human blood incorporates a cascade of activation enzymes and proenzymes. The removal or change of any single protein in the blood clotting cascade frequently has catastrophic results as in hemophilia.

Irreducibly complex systems pose a difficulty for Darwinian evolution, that is random variation of inherited characteristics through mutation of the genetic code and natural selection, if the parts of the system are independently coded. The creation or evolution of such systems apparently requires at least two simultaneous mutations. Any single mutation will break the system. At least two parts must change together. Since the parts are encoded independently, at least two mutations must occur simultaneously. Further, since the parts are encoded independently and the mutations are presumably random, the vast majority of multiple simultaneous mutations also break the system.

In the case of body plans, the genetic code is not known. Various regulatory genes and regulatory systems have been proposed. It seems likely that biomechanical parts such as bones are coded in an interdependent, highly correlated fashion. Various researchers and commentators have suggested that a single mutation or a small number of mutations in the hypothetical regulatory genes governing the development of embryos could generate large-scale changes such as a change in the number of vertebrae in the spinal column[2,3]. In particular, this explanation has been offered to explain the apparent proliferation of new body plans during the Cambrian explosion.

In biochemical systems most proteins are currently believed to be independently coded. The textbook rule is one gene for each enzyme. Several exceptions are known. These include biochemical pleiotropy where a single gene codes for a sequence of amino acids that is incorporated in several different longer proteins. Overlapping genes have been discovered. Some proteins are synthesized by cells and then subsequently broken down into smaller child proteins that are reused for other purposes. An analogy can be made to a tool that is used for one purpose and disassembled into smaller tools used for other purposes. These cases are frequently treated as exceptions to the general rule that proteins are independently coded.

The problem of irreducible complexity in Darwinian evolution can be resolved if the living systems are not irreducibly complex. No rigorous method of proving that a system is irreducibly complex exists. For example, to prove that a system with two parts is irreducibly complex requires an exhaustive search of all possible precursors where only one part has been changed. While it is usually easy to prove that the removal of any single part will break the system, it can be quite difficult to prove that there is no possible single change of a single part within the system that also yields a working system.

Irreducible complexity is a consequence of the tight tolerance between the parts. In the simple example of the lock and key both the lock and key can be modified very slightly within a mechanical tolerance and still function. Some simple locks and keys have a very broad tolerance and can, for example, be opened with a crude lock-pick such as a wire. Many proteins appear to function with a very tight tolerance so that only one or a few other proteins react with the protein. The tight tolerances in biochemical systems appear necessary for function but this is difficult to prove.

A seemingly irreducibly complex system might have evolved from a precursor system of parts with very loose tolerances. Natural selection slowly produced the tight tolerances observed in the existing biochemical systems such as the blood clotting cascade. A single mutation would reduce the tolerance of single part slightly, enough to improve the system and yet still function with the other parts. The primary obstacle to this explanation is that many systems appear to require a tight tolerance to function at all. This explanation is a special case of "the system is not irreducibly complex" explanation.

Irreducibly complex systems can be explained if a part or group of parts evolved through Darwinian evolution as parts of another system that is not irreducibly complex, a reducibly complex system. A single chance mutation might then combine the parts from two or more other reducible systems into a single irreducibly complex system. This is probably the most commonly cited explanation for irreducibly complex biomechanical and biochemical systems. Experience with both living systems and man-made machines illustrates that random recombinations of parts almost always result in non-functional systems. Most man-made machines made of several parts and most living systems are networks of parts where each part is tightly matched to one or a small number of other parts. The other parts are physically adjacent in mechanical systems. In biochemical systems, the other parts have tightly matched chemical binding sites that insure that only the few associated parts interact even though hundreds or thousands of different parts float free in a water-based solution. In other words, each part has specific shared interfaces with a handful of other parts. Consequently, random recombinations of parts or groups of parts nearly always result in combinations of incompatible parts. Thus this explanation involves an appeal to an unlikely coincidence not unlike several parts changing in tandem.

The problem of irreducible complexity can also be resolved if the parts of the system are not independently coded or were not independently coded when the system evolved. A simple toy, the jigsaw puzzle, illustrates this. A jigsaw puzzle meets the definition of irreducible complexity presented above. Removal of any single piece or modification of any single piece breaks the puzzle. If the pieces of the puzzle are independently coded – for example as an ordered list of the vertices of the piece – then any single mutation and the vast majority of simultaneous multiple mutations breaks the puzzle. This independent coding of pieces is not how jigsaw puzzles are designed or created. Jigsaw puzzles are formed by cutting a parent piece into a set of matching pieces with a literal or figurative jigsaw. If the jigsaw puzzle is coded as the shape of the parent piece, for example an ordered list of the vertices of the parent piece, and the rules for cutting the parent piece, any single mutation will produce a valid irreducibly complex jigsaw puzzle. A single mutation will change several pieces simultaneously in a correlated fashion.

The two ways of coding a jigsaw puzzle are examples of representations. A representation is a way of describing a system, usually exactly. In general, there are many representations for a system. Although different representations describe the same system, sometimes a problem will be intractable in one representation and easy to solve in another representation. For complex systems of parts, there is a family of part-level representations in which each part is described independently. Part-level description or part-level code may be substituted for part-level representation. There is also a family of system-level representations in which parts are described in an interdependent fashion. A system-level representation will preferentially or exclusively represent parts that are correlated in some essential way necessary for the function of the system. System-level description or system-level code may be substituted for system-level representation. In the following discussion, it is important to realize that the system-level representation and the part-level representations are equivalent. The system-level representation can be transformed into the part-level representation and vice-versa. This equivalence of two representations is sometimes called duality by mathematicians and physicists. It will be argued that it is often easier to design systems and solve certain problems in a system-level representation. However, efficient mass-production is often performed in a part-level representation.

A system-level genetic code is a hypothetical genetic code that preferentially or exclusively represents systems of tightly coadapted parts. In systems designed by humans, this typically involves coding the interface shared by two or more parts as a single element of the code and coding the properties of the parts that are independent of the interfaces as independent elements of the code[4]. Some properties are shared by all parts. Some properties are shared by groups of parts. Some properties are unique to a single part. For example, a jigsaw puzzle can be made of wood or cardboard. This is usually a shared property of all pieces in the puzzle. Ideally, the code for the interface elements is defined so that any string of symbols codes a valid interface. Any single-step mutation of the interface element generates another valid interface element. A mutation of the interface element can simultaneously change two or more parts in the system. Similarly, a single mutation of a property shared by all parts will change all parts in the system simultaneously. A system-level genetic code combined with random variation and natural selection could evolve irreducibly complex systems.

Several reasons for suspecting a system-level genetic code exist[5]. The fossil record contains numerous cases of the sudden appearance of new forms. Intermediate forms are frequently absent or unrecognized. There is a troubling lack of convincing intermediate forms between seemingly widely separated forms where many, many intermediates would be expected. The fossil record looks suspiciously like large-scale changes, saltations or systemic macromutations that would require several parts to change at once, have occurred on several occasions. For example, in the Cambrian explosion of about 600 million years ago, most of the invertebrate phyla appeared in the fossil record during a period of only 50 million years, possibly 10 million years by some estimates. Intermediates and plausible precursors to the different invertebrate phyla are absent. The Cambrian explosion suggests a discontinuous jump from primitive multi-cellular organisms to fully functional animals.

The transition from single-celled organisms to functional multi-cellular organisms required the creation of a biochemical system that forced the cells to cooperate. Amongst other things this system inhibits the cells from devouring one another. At the same time, the cells must recognize and attack foreign cells. Otherwise the foreign cells will quickly devour the defenseless multi-cellular organism. It must insure that nutrients are distributed from the surface of the organism to all interior cells. Otherwise the interior cells will die. It is not sufficient to prevent a cell from directly devouring its fellow cells. The multi-cellular control system must insure that no cell in the organism consumes more than its fair share of the nutrients and grows out of control, starving the other cells or poisoning the other cells with waste products. Conversely the multi-cellular organism must replace cells that die. Unlike single-celled organisms, multi-cellular organisms inherently require the controlled growth of cells. Uncontrolled, exponential growth is deadly to the organism as in cancer. The system must insure that any waste products are expelled from the organism. Otherwise the waste products will build up and kill the multi-cellular organism. The system must also insure that the cells in the organism physically stick together. Some of these features might have been built up in a slow, gradual manner through a gradual transition from single-celled organisms to mats of cells such as the mats of cyanobacteria that produce stromatolites. In a mat every cell has access to the primordial seas providing nutrients and waste disposal. Inhibition of cannibalism might be sufficient without a complex feedback system to control cellular growth. However, the transition to a functional multi-cellular organism with interior cells seems to be a large functional leap exactly as the fossil record seems to indicate. Similarly, the transition from simple multi-cellular organisms to animals seems to be a large functional leap. Intermediates may not be possible.

Traditionally the absence of intermediates has been attributed to the imperfection of the fossil record or to inadequate search of the fossil record. More recently, the theory of punctuated equilibrium in which evolution occurs rapidly in small isolated populations has been proposed[6]. Neither explanation is a convincing explanation for the marked absence of intermediates between phyla. More intermediates would be expected between widely separated forms such as different phyla than between closely related species. With an imperfect fossil record or inadequate search of the fossil record, there should be more examples of intermediates between widely separated species such as species belonging to different phyla than between closely related species. While some examples of possible intermediates between widely separated forms such as the *Archaeopteryx* exist, in general intermediates between widely separated forms are rarer than possible intermediates between closely related forms. Similarly, while punctuated equilibrium theories can account for an absence of intermediates between closely related species, such as the dog and the fox, this is not a plausible explanation for gaps between widely separated species unless systemic macromutations are incorporated in the theory. Long periods of stasis in which the fossilized parts of species undergo no change are positively documented.

Systemic macromutations or saltations, the "hopeful monsters" of Richard Goldschmidt, are implausible without new physical phenomena or an intelligent agent if the parts of the living systems are independently coded[7]. The jump from primitive multi-cellular organisms to multi-cellular invertebrates that the Cambrian explosion suggests would require many simultaneous harmonious changes. A jump from a terrestrial rodent to a functional bat would require dozens of bones and muscles to change simultaneously. If the parts are independently coded then the probability of random variation producing these jumps is essentially zero.

A system-level genetic code permits a single-step mutation to change two or more parts harmoniously and can, in principle, leap major functional gaps, such as the gaps between phyla, in a single step. In the toy example of the jigsaw puzzle the addition of a new cutting rule might correspond to a major jump between widely separated species. A change in a pre-existing cutting rule might correspond to a small jump between closely related species. Although a mutation in a system-level genetic code always generates a system of coadapted parts, most of these systems will be negative or neutral mutations. A species may remain stable for a long time.

Major gaps such as those between phyla will rarely be spanned. If a major jump occurs, smaller more probable jumps will then populate the branch of the tree of life corresponding to the new phylum. Once a single species of a new phylum is created, there will be a rapid, in geological time, proliferation of species belonging to the new phylum. Thus periods of mass proliferation of new species such as the Cambrian explosion are expected immediately following a major leap across a functional gap – for example, the leap from primitive multi-cellular organisms to animals. Once a working example of a basic type appears, this example will diversify into many new species within that type much faster because the gaps are smaller and the probability of spanning a small gap is greater than spanning a large gap. Thus the major features of the fossil record, both stasis and the sudden appearance of new species and entire types, can be explained with a system-level genetic code.

This model requires that the creation of new organs or complex biochemical systems precedes the new species becoming reproductively isolated from the parent species in organisms that reproduce sexually. The first instance of the systemic macromutation needs to mate with members of the parent species to reproduce. For example, a systemic macromutation creates a functional proto-bat from a species of terrestrial rodent. The proto-bat mates with the rodents producing a sub-population of flying, winged rodents. Over time this population becomes isolated from the parent population. The bats nest in trees or other high places inaccessible to the ground rodents. Additional changes such as the development of bat sonar produce a reproductively isolated population of bats, a new species.

Several candidates for irreducibly complex biochemical systems exist. These include the blood clotting cascade, the cilium, the bacterial flagella, and the immune system. These biochemical systems suggest the simultaneous or nearly simultaneous appearance of several coadapted proteins.

Gene knockout studies in which a single coding gene is removed usually show that a given gene affects several different systems in apparently disjoint ways. The gene does not change all of the systems in a harmonious manner as a system-level genetic code mutation would - at least sometimes. Rather all of the systems that appear to use the protein coded by the gene break. This substantially reduces the likelihood of the largely hypothetical positive mutations presumed to drive evolution. Since a protein is frequently reused in multiple different systems or used to form several longer proteins used in different systems, in general a mutation must not only improve the protein for one system but preserve or improve its function in all systems that utilize the protein. *Or* all of the different systems that use the protein must change together when the constituent protein changes. These results also suggest a system-level genetic code.

The pattern of differences in the amino acid sequences in proteins and nucleotide base pair sequences in the DNA between different species, usually explained through the molecular clock hypothesis, is difficult to account for with the traditional part-level genetic code. Specifically, different species appear to have widely differing generation times and annual rates of mutation. The molecular clock hypothesis appears to require a constant rate of mutation per unit time across hundreds of very different species. The rates must also differ from protein to protein because some proteins such as cytochrome C have much wider variation across species than other proteins such as histone. Systemic macromutations of the biochemical system from mutations in a system-level genetic code (or other mechanisms) would arguably cause wide variations between all parts when a large jump occurs in the system and smaller variations between parts when a small jump occurs. This would reproduce the typological pattern often attributed to the hypothetical molecular clock.

The observed genetic code contains features that suggest a system-level genetic code. The genetic code for the blood clotting cascade contains similar sequences in different genes and within the same gene. These frequently appear to code for sequences in the proteins that bind to Vitamin K, another constituent of the blood clotting cascade[8,9]. This may be an example of a reusable standard interface component such as a connector. There are also adjacent pseudo-genes that appear to be non-functional copies or near copies of the coding genes. Pseudo-genes are a common part of the genetic structure. This may be similar to a master copy used to fabricate production dies but never used directly to manufacture the parts. This analogy would reverse the presumed order of gene duplication with the pseudogene acting as a precursor to the coding gene. The gene duplication transforms the non-coding pseudo-gene into the coding gene, several closely related coding genes, or homologous sequences within several coding genes. These correlated structures in proteins are usually attributed to random gene duplication and random shuffling of regions delimited by the introns. Regardless of their specific interpretation, correlations between seemingly independent parts would be expected if a system-level genetic code exists.

A system-level genetic code governing the morphology of living systems could resolve a major problem with homology. Homologous organs such as eyes frequently follow substantially different formation pathways during the development of the embryo[10]. The same organ or homologous structure may develop from completely different precursor structures in different species. This is extremely difficult to explain if the different parts such as bones are coded independently. This unstated assumption led early advocates of evolution such as Charles Darwin and Ernst Haeckel to expect similar development of embryos and homologous organs in different species. Similarly Haeckel's erroneous claim that "ontogeny recapitulates phylogeny" is a logical deduction if one assumes that each physical part of the adult organism is coded independently and therefore must develop essentially independently in the embryo. However, consider the toy example of the jigsaw puzzle coded as a series of rules for cutting a parent piece into matching smaller pieces. The order in which the puzzle is cut can be varied by swapping different cutting stages. If the order in which the cutting operations are executed varies, identical jigsaw puzzles can be produced by substantially different pathways.

Human beings frequently design and manufacture irreducibly complex systems. Human beings frequently use system-level representations to design and manufacture these systems. Design, engineering, and manufacturing may provide the best insight into how natural system could contain and implement a system-level genetic code. The human example is explored in detail below. Simple examples of how biochemistry might implement similar methods are suggested. Detailed examples of system-level codes are presented.

2. THE TEMPLATE MECHANISM

Living systems contain numerous examples of tightly matched, coadapted parts, both physical parts and biochemical parts. While such parts appear rare in nature outside of life, all but the simplest man-made machines use such parts. How then do humans routinely make such parts and how might biochemical systems emulate this?

Probably the most common method to produce tightly matched parts is to use one part as a template for a second matching part. In mechanical systems, the first part is manufactured. The second part is machined or molded to match the first part at a common boundary.

A signal encoder and decoder pair is an example of an irreducibly complex system of two parts. Signal encoders and decoders are used for a variety of purposes. Signal encoding is used to transmit signals, to encrypt signals, to protect signals from transmission errors, and to compress signals for transmission over bandwidth limited channels or data storage. A signal encoder without a decoder is useless. A decoder without an encoder is useless. In many cases, the encoder and decoder are tightly matched. A small change or error in either the encoder or decoder will break the system. The decoded signal is unrecognizable.

A television transmitter and receiver is a familiar example of a matched signal encoder and decoder. A European PAL television receiver cannot display an American NTSC television signal. In general, a small change in the television transmitter or receiver alone results in an unrecognizable television signal. This is a general feature of many analog and digital communications systems.

In general, the signal encoder is designed and manufactured first. A decoder needs an encoded signal. The encoder naturally comes first. In most cases there is a fixed mathematical relationship between the encoding and decoding steps. For example, the encoder multiplies by a matrix and the decoder must multiply by the inverse of that matrix. The inverse matrix can be derived from the forward matrix in the encoder. The encoder acts as a template for the decoder.

In biochemical systems, a special enzyme or system of enzymes might be able to build a matching protein from a template protein. Since separate genes coding for matched proteins are frequently observed, it would seem that some mechanism to map the protein back into the DNA sequence would be required.

Alternatively, an algorithm for deriving the DNA sequence of the matching protein from the template protein's DNA sequence may exist. This does not require a complete understanding of protein folding – the ability to predict the structure of the protein from the DNA sequence. All that is needed is an invariant transform between the sequence producing the topology of the folded protein and a sequence producing a matched topology at least for a large class of proteins used in living systems. For example, replacing the negative amino acids with positive amino acids and vice versa while leaving the hydrophilic and hydrophobic amino acids alone might be sufficient to create a complementary protein for some classes of proteins. In this case, one gene would be derived from the other seemingly independent gene. It is likely that some process, usually interpreted as random, can copy genes at times, creating the common non-functional pseudo-genes. Perhaps a more sophisticated process creates one gene from another. In this case, the genetic code would resemble:

*(DNA sequence for a part)(DNA sequence instructing the system to make a matching part from the sequence for the preceding part)(DNA sequence for the matching part)*

A more general mechanism would subdivide the parts, the sequences for the proteins, into interface domains within the proteins and non-interface domains. For example, a chain of interacting parts might be coded as:

*(START)(DNA sequence for a non-interface domain)(DNA sequence for an interface domain)(STOP)(DNA sequence instructing the system to make a matching interface domain from the preceding domain)(START)(DNA sequence for the matching interface domain)(DNA sequence for a non-interface domain)(DNA sequence for an interface domain)(STOP)(DNA sequence instructing the system to make a matching interface*

*domain from the preceding domain)(START)(DNA sequence for the matching interface domain)...*

This simple model bears a superficial resemblance to the common descriptions of the mysterious non-coding introns in eukaryotic cells if the introns are interpreted as delimiters between functional domains in proteins such as interfaces. The introns appear to be used in exactly this way in the immune system. Other interpretations of the introns are possible within the general theory of a system-level genetic code. For example, no mechanism for deriving the matching interface domain from the template domain may exist. However, the introns may delimit pairs of matching interfaces that are copied or changed together during directed mutation events possibly governed by a system-level genetic code.

These ideas are elaborated further in Section Four on Standards and Reusable Interfaces.

3.  THE JIGSAW MECHANISM

The other common mechanism for producing systems of coadapted parts is to subdivide a larger piece into two or more matching pieces through bisection, subdivision, or fragmentation[11]. In a mechanical system, the analogy of the jigsaw puzzle is exact. In the jigsaw puzzle, a larger piece is cut repeatedly by a literal or figurative jigsaw to produce a large number of tightly coadapted pieces.

In top-down design a system is represented as a single block initially. The block is then iteratively subdivided into smaller blocks with well-defined interfaces. The designer may either define an interface between the blocks during the subdivision process or use a standard interface. The subdivision is repeated until the blocks have a simple one-to-one relationship with simple parts. At this point either custom parts are designed and fabricated or standard parts are used. Top-down design is a more abstract example of a jigsaw mechanism.

A possible biochemical analog of the jigsaw mechanism would be a chemical process, for example hydrolysis, or physical process that breaks a long biopolymer such as a protein or RNA strand into smaller pieces. In particular, long proteins fold into complex three dimensional structures in which sequences with complementary shapes and chemical affinities are adjacent. A physical or chemical jigsaw mechanism might be able to break apart the long protein into highly correlated, frequently coadapted pieces. The jigsaw mechanism would cleave open loops that did not exhibit a chemical affinity.

In this model, initially a biochemical system of several coadapted parts would be coded as the sequence for a long parent protein and the sequence for a jigsaw protein that cuts the parent protein into pieces. Since protein folding is complex, the cutting operation would not always produce a system of coadapted pieces. However, this would occur sometimes. A single mutation in either the code for the parent protein or the jigsaw protein would modify the entire system, not a single part.

*(DNA sequence for the parent protein)(DNA sequence for the jigsaw protein)*

In this representation, the complex system of biochemical parts could evolve more rapidly and plausibly than in a part-level representation.

This mechanism could be applied recursively to represent extremely large, complex, hierarchical networks of parts. The jigsaw protein could cut the parent protein into several parts including a secondary jigsaw protein and a secondary parent protein. Conceivably, the process could be iterated indefinitely to represent very complex systems.

*(DNA sequence for primary parent protein)(DNA sequence for the primary jigsaw protein)*

The action of the primary jigsaw protein would create a collection of parts:

*(secondary parent protein)(secondary jigsaw protein)(protein 3)(protein 4)...(protein N)*

The action of the secondary jigsaw protein would create a collection of parts:

*(protein 1.1)(protein 1.2)...(protein 1.m)(protein 3)(protein 4)...(protein N)*

Thus very complex networks of proteins could be represented by the jigsaw mechanism. Additional complexity could be achieved by using multiple jigsaw proteins, corresponding to multiple cutting rules in the toy example of a standard jigsaw puzzle:

*(DNA sequence for parent protein)(DNA sequence for jigsaw protein 1)(DNA sequence for jigsaw protein 2)*

The system makes the parent protein, uses jigsaw protein 1 to chop up the parent protein and then uses jigsaw protein 2 to further fragment the parts produced by the first pass.

Over time mutations would introduce stop codons into the DNA sequence for the parent protein. If these stop codons appeared in locations different from the locations where the jigsaw protein cuts the parent protein, these stop codons would clearly break the system and be eliminated by natural selection. However, sometimes a stop codon would appear in the same location in the sequence as the location where the jigsaw protein cuts. This is illustrated as:

*(DNA sequence for Part One)(DNA sequence for the rest of the parent protein)(DNA sequence for the jigsaw protein)*

In this case, the system would continue to function. Producing a system of coadapted parts by cutting a larger protein into pieces is probably slower, more cumbersome, consumes more energy, and possibly more error-prone than producing a system of coadapted parts from separate genes. Thus, over time, natural selection would favor the variants with stop codons at the locations in the sequence corresponding to the locations where the jigsaw protein cuts. Eventually the genetic code for a system with $N$ parts would become:

*(DNA sequence for Part One)(DNA sequence for Part Two)...(DNA sequence for Part N)(DNA sequence for the jigsaw protein)*

The jigsaw protein would become unnecessary. The now independent genes could drift apart in the genetic code for the organism, obscuring the initial relationship. Once the separate genes formed, a period of stasis would occur where it was very difficult for the complex system to evolve since single-step mutations could no longer change the parts together.

New complex biochemical systems could be generated by a "white noise" mutation event that generated a long sequence of random base pairs coding a new long parent protein at an appropriate location in the genetic code relative to the sequence for a jigsaw protein. The "white noise" mutation event might occur during cell division. Conceivably the jigsaw gene could remain dormant until an appropriate parent protein appeared.

Often the "white noise" mutation event would generate a complex system of proteins that conferred no advantage on the organism. These mutations would be eliminated by natural selection. Occasionally the "white noise" mutation event would create a complex system of proteins such as the early blood clotting cascade that conferred an advantage and represents a systemic macromutation. For a while, the new system could evolve rapidly since the parts were coded in an interrelated manner. Eventually mutations introducing stop codons at the appropriate locations in the sequence decouple the biochemical parts of the system. The system freezes into stasis and no longer evolves significantly. This model thus reproduces the rapid flowering of new types of life followed by long periods of stasis seemingly observed in the fossil record.

A more radical model is that the genetic code maintains two representations of the complex biochemical systems at all times. One representation is the part-level genetic code, the coding genes that are known, and the other representation is a system-level genetic code as described above. In this model, the complete code for a complex biochemical system with *N* parts would be:

*(DNA sequence for the parent protein)(DNA sequence for the jigsaw protein)(DNA sequence instructing the system to derive the sequences for a series of parts from the preceding jigsaw mechanism system-level sequences)(DNA sequence for Part One)(DNA sequence for Part Two)...(DNA sequence for Part N)*

Some unknown biochemical process, possibly during cell division, derives the part-level code from the system-level code. Note that the DNA sequence for the parent protein contains the sequences for each of the constituent parts without the start and stop codons found in the coding genes. This bears a resemblance to common descriptions of the pseudogenes that are often found near coding genes.

4. STANDARDS AND REUSABLE INTERFACES

Once an interface exists, humans frequently reuse the interface in many different parts and machines. Millions of houses use the same standard wall plug. The same standard credit card size and shape is used by Automatic Teller Machine (ATM) and credit card machines throughout much of the world. Once an interface exists, it can be standardized and used to create a large variety of parts and machines that fit together.

Among humans a standard is usually established by a standard-setter such as a government, company, or other organization. For example, the standard size and shape of credit cards is a standard from the International Organization for Standardization (ISO), a quasi-governmental organization affiliated with the United Nations. From a technical point of view, a standard usually consists of a detailed written specification, a standard document, and one or more working prototypes that are used as master copies. In the standards field, the working prototypes are referred to by a variety of names such as reference design, test model, verification model, simulation model, and so forth. In principle all working implementations of the standard are manufactured to match the behavior of the working prototype.

The working prototypes often differ from the mass-produced goods or services that conform to the standard. For example, the standard pounds and kilograms maintained at the National Physical Laboratory in England are made of platinum. The clock used to provide the standard Greenwich Mean Time is a highly accurate atomic clock. The prototypes are frequently designed to be extremely durable, to be more accurate than the mass-produced goods and services, and often show direct evidence of the template or jigsaw processes used to insure matching interfaces.

In mass production, the working prototypes that serve as master copies are often the point of transition from a system-level representation to a part-level transition. The system is designed using a system-level representation to insure that the many parts of the system work together. Once a working prototype exists, a part-level representation is frequently used to mass-produce the goods or services. For example, individual parts may be mass produced using separate production dies for each part. These dies show no obvious signs that one part was formed by making a mold of the matching part. In some systems, such as Videocassette Recorders (VCR)'s, some matching parts, the videotapes, must be manufactured independently of the other parts, the recorders. The close interrelationship between the parts is disguised by the mass production methods.

The working prototypes that form the basis of most standards are frequently invisible to most end-users and even many designers and engineers. These prototypes are used only infrequently to insure that the factories, production dies, and so forth conform to the standard. A business or organization that is not itself the standard-setter may create secondary master prototypes derived from the original master prototypes maintained by the standard-setter. An entire cascade of masters leading from the original master, the "gold" master or "gold" standard, maintained by the standard-setter to a production die used on the factory floor may exist.

Standards established by humans often incorporate other features that may cast light on how similar systems in biochemistry might be implemented. Standards frequently have a name and logo that is protected through trademark and service mark laws. Typically a standard-setter allows or requires a good or service to use the standard name and logo only if the good or service conforms to the standard as specified by the standard-setter.

The standard-setter may demand a license fee to use the trademarked name and logo. The standard-setter may reserve the use of the name and logo for goods and services that it manufactures. The standard is a proprietary standard.

Names and logos are used for marketing and business purposes that seem to have no analogy in living systems. However, the names and logos serve a practical purpose that might be emulated in living systems. Specifically, the names and logos provide easily accessible labeling of coadapted parts. For example, most Video Cassette Recorders (VCR) conform to the Video Home System (VHS) standard. VCR's bear the VHS name and logo. Compatible video tapes also bear the VHS name and logo. This enables customers to quickly identify the compatible parts and avoid attempting to combine incompatible parts which can have disastrous consequences.

In biochemical systems, it might be useful or even necessary to have a tag that identified parts such as regions of proteins or genes as conforming to a biochemical standard. The tag sequence itself, much like a logo, might serve no direct identifiable purpose and yet be repeated across many different parts, that is genes or proteins. Its sole purpose would be to identify compatible parts to the cellular machinery. Some of the repeated or nearly repeated sequences in proteins and genes might serve this purpose.

Standard-setters frequently employ various forms of intellectual property including trademarks, copyrights, patents, and trade secrets to maintain control over the standard. This is frequently used to establish monopolies on goods or services conforming to the standard. A monopoly for profit seems to have no analog within a biochemical system. However, a mechanism for insuring exclusive centralized control over a standard does serve a practical purpose unrelated to profit. In the absence of centralized control, a single standard-setter, multiple competing, mutually incompatible versions of the standard can develop. In this situation, the primary advantage of the standard is lost.

In biochemical systems, some mechanism of insuring that all parts, that is proteins, using a standard interface contain interfaces conforming to a single version of the standard should exist. All interfaces should be required to conform to a single prototype or master version of the interface somewhere within the genetic code. The standardized interfaces should be identical to within some tolerances established by the system. Non-conforming interfaces should be eliminated by some enforcement mechanism. In particular the system needs to protect against duplication of the master copy followed by mutation of the duplicate resulting in two incompatible overlapping standards.

If standards exist in the biochemical system, one would see the repeated reuse of standard amino acid sequences in proteins and quite possibly multiple copies or near copies of the genes coding for proteins. These extra, rarely used copies of genes would be analogous to the master prototypes used in standardized manufacturing. The coding genes, corresponding to production dies, would be derived from the master copies.

4.1  A SYSTEM-LEVEL CODE FOR CHAINS OF MATCHED PARTS

Once standard interface components exist, it is possible to code for complex systems in such a way that all single mutations generate a complex system of co-adapted parts. For example, let the upper-case letter *A-Z* represent the standard interfaces. The upper-case letter *A* represents a complementary pair of components *A-* and *A+* forming a standard interface. Examples of *A-* and *A+* are a lock and key, a plug and jack, a telephone transmitter and receiver, and so forth. Let the lower-case letters *a-z* represent components that play no role in the interface between parts.

| Symbol in Code | Meaning | Terminal Symbol |
| --- | --- | --- |
| (whitespace) | Separates Discrete Parts | Terminal |
| *A-, B-, C-, ...* | Negative Half of Interface | Terminal |
| *A+, B+, C+, ...* | Positive Half of Interface | Terminal |
| *a,b,c,...* | Attribute of a Part | Terminal |
| *A,B,C ....* | Standard Interface | Non-Terminal |

This system-level code is a special case of a context free grammar. A context free grammar (CFG) is a concept used in compilers for computer languages and in linguistics. Compilers are programs that convert a program written in a high-level computer language such as C, C++, or FORTRAN to binary machine language instructions that a computer understands[12]. A context free grammar consists of:

(1) A finite terminal vocabulary $V_t$

(2) A finite set of different, intermediate symbols, called the nonterminal vocabulary $V_n$

(3) A start symbol $S \in V_n$ that starts all derivations. A start symbol is sometimes called a goal symbol. The start symbol is a member of the set of nonterminal symbols.

(4) $P$, a finite set of productions, sometimes called production or rewriting rules, of the form $A \to X_1 \ldots X_m$ where $A \in V_n$, $X_i \in V_n \cup V_t$, $1 \leq i \leq m, m \geq 0$

The context free grammar starts with the start symbol. The production rules are applied until a series of only terminal symbols are reached. One can think of the production rules and the set of nonterminal symbols, the nonterminal vocabulary, as the system-level code. The final sequence of terminal symbols is the part-level code. In a hypothetical system-level genetic code the terminal symbols would probably correspond to functional domains within coding genes.

In the context free grammar for this simple example system-level genetic code, the production rules for the context free grammar specify a unique message rather than an allowed syntax of a family of messages. The message is encoded as a series of production rules for rewriting the system level code. These rules are iterated until a message formed of terminal symbols is reached.

This simple example code has a single rewriting rule that translates a standard interface such as *A* into *A- A+* where *A-* and *A+* are separated by whitespace.

A chain or cascade of interacting parts can be represented as a series of the letters. For example:

*aAabBaBcCa*  (System-Level Code for a Chain of Five Coadapted Parts)

This is equivalent to the discrete parts:

*aA-   A+abB- B+aB- B+cC- C+a*   (Part-Level Code for a Chain of Five Coadapted Parts)

The system-level code acts as the master copy. The code translates the interface components *A-Z* into the two complementary parts: *A-* and *A+, B-* and *B+*, and so forth. Spaces delimit discrete parts in the part-level code and might correspond to start and stop codons in the DNA genetic code.

In the language of context fee grammars, the code is:

(system) : *aAabBaBcCa*

*A : A- A+*

*B : B- B+*

*C: C- C+*

Any single mutation in the system-level code such as addition, deletion, or change of any single letter to another letter in the alphabet yields another system of coadapted parts. For example, if the first upper-case *B* changes to a lower-case *b*, the code becomes:

*aAabbaBcCa*  (System-Level Code for a Chain of Four Coadapted Parts)

This is equivalent to the discrete parts:

*aA- A+abbaB- B+cC- C+a* (Part-Level Code for a Chain of Four Coadapted Parts)

Any single mutation of the interfaces such as *A-* and *A+* in the part-level code will break the chain. There is also no way for a single mutation in the part-level code to bisect a piece into two complementary pieces preserving a chain of interacting parts. For example, *A+abbaB-* cannot be converted to *A+aB- B+baB-* by a single mutation in the part-level code.

The simple example allows some hard numbers to illustrate the difference between a system-level and part-level code. Consider the simplest system of coadapted parts, a system of two complementary parts:

*aAb* (System-Level Code for Two Coadapted Parts)

*aA- A+b* (Part-Level Code for Two Coadapted Parts)

Ignoring the two ends *a* and *b* for simplicity, there are twenty-five (25) single mutations of *A* yielding valid systems, for example *A -> B*. There are twenty-six (26) single mutations of *A* that yield invalid systems, for example *A -> b*. However, in the part-level code, there is no single mutation of the interface pieces *A-* or *A+* that yields a valid system. There are only twenty-five (25) double mutations that yield valid systems. There are 26*26 plus 25*24, a total of 1276, double mutations that break the system and yield invalid systems where the two parts will not work together. The probability of random variation in the part-level code improving the system is very small whereas the probability in the system-level code is actually quite high.

If the probability of a single mutation is one (1) per one-thousand (1000) generations, then the probability of a double mutation is one (1) per one-million (1,000,000) generations. Just as an example, suppose only one change represents a superior system, for example *aBb* also represented as *aB- B+b* is better than *aAb*. Organisms that contain the system coded by *aBb* are more able to survive and reproduce than those with *aAb*. For example, if *aAb* and *aBb* are parts of a primitive blood clotting cascade, then *aBb* might be less prone to accidental activation by proteins in the plaque forming on the inner surface of blood vessels, a speculated cause of heart failure. In the system-level code it would take about 26,000 generations to improve the species. In the part-level code it would take 1,000,000 generations on average simply to get one pair of simultaneous mutations. Of these, only 25 out of about 1301 double mutations are even valid systems. And only one (1) is the better case. It would take about 650,000,000 generations to make the one improvement, to find the better *aBb* case. The system-level code evolves about 20,000 times faster than the part-level code.

The single mutation rate of one per one-thousand generations must be significantly higher than the actual observed mutation rate for proteins in the blood clotting cascade. If this were true, one in a thousand children would be born with usually deadly blood disorders. The incidence of hemophilia and other congenital blood diseases would limit the rate to something more like 1 in 3,500 generations even if all cases of hemophilia and other diseases were attributed to a current mutation instead of one inherited from past generations.

The *aAabBaBcCa* system-level genetic code corresponds to a rigorous top-down design in which a designer partitions a system into black boxes. Standard interfaces are selected and used to specify the relationship between the different black boxes. For example, a stereo system designer selects the standard RCA stereo jack and plug as the interface between the different modules in the stereo system such as receiver, amplifier, and speakers. The standard interfaces such as the RCA jack and plug correspond to the upper-case letters *A-Z* in the code. Each black box in the design has attributes that are independent of the interfaces specified. These attributes can be changed without modifying the interfaces. In a stereo this might correspond to the choice of a vacuum tube based FM tuner or a transistor-based tuner. These attributes correspond to the lower-case letters *a-z* in the code. For example, *a* might code for a vacuum tube based FM tuner. The lower-case letter *b* might code for a transistor-based FM tuner.

The human designer achieves great efficiency by choosing the necessary interfaces for the parts from a library of pre-existing interfaces, coded by *A-Z,* and a library of pre-existing attributes, coded by *a-z*.

Naively extrapolated to DNA and proteins, the system-level code *aAabbaBcCa* might correspond to the pseudo-genes and other "junk" DNA and the part-level code *aA-A+abbaB- B+cC- C+a* might correspond to the coding genes for a cascade of coadapted proteins. In this scheme, the coding genes would be derived from the supposedly non-coding DNA by a highly organized process that might superficially resemble random duplication of genes and shuffling of parts of genes. The mysterious non-coding introns might be interpreted as markers and spacers separating interfaces such as *A+* from non-interface sequences such as *abba*. Complex networks of coadapted proteins such as the blood clotting cascade would clearly require a more sophisticated system-level code than this simple example.

The first living organisms might have contained a simple version of a system-level genetic code using one or a few simple standard interfaces between the parts – the initial complex biopolymers such as RNA or proteins. For example, Lego™ blocks use a single standard interface between parts and yet can be assembled into a wide variety of different systems. This would have permitted rapid evolution of complex systems including more complex versions of the system-level genetic code.

4.2 A SYSTEM-LEVEL CODE FOR BRANCHING NETWORKS OF PARTS

Machines and living organisms contain far more complex networks of interacting parts than linear chains of interacting parts. The blood clotting cascade contains complex feedback loops that control blood clotting. More sophisticated system-level codes than the simple example above are needed.

Expressed in English the system-level code *aAabBbaCd* is "The first part, which has attribute *a,* uses the interface *A* to activate the second part, which has attributes *ab,* which uses the interface *B* to activate the third part, which has attributes *ba,* which uses the interface *C* to activate the fourth part, which has attribute *d*." In English much more complex networks of interacting parts can be described by a linear chain of symbols. For example, "The first part, which has attribute *a,* uses the interface A to activate the second part, which has attributes *ab,* which uses the interface *B* to activate the third part, which has attributes *ab,* and the interface *C* to activate the fourth part which has attribute *d*".

*aAabBab&Cd*  (System Level Code for a Branched Network of Four Interacting Parts)

*aA- A+abB-C- B+ab C+d* (Part Level Code for a Branched Network of Four Interacting Parts)

A new non-terminal symbol *&*, in *&C*, is introduced to encapsulate "and uses interface C to activate" in the English description.

*aAab&Bab&Cd* means "The first part, which has attribute *a,* uses interface *A* to activate the second part, which has attributes *ab,* and uses interface *B* to activate the third part, which has attributes *ab,* and uses interface *C* to activate the fourth part, which has attribute *d*". The part-level code is:

*aA-B-C- A+ab B+ab C+d*

Additional symbols functioning as open and close parentheses are needed to implement complex hierarchies of branching networks. For example:

*aAb(BcCd)&DeEf* means "The first part, which has attribute *a,* uses the interface *A* to activate the second part, which has attribute *b* and which uses interface *B* to activate the third part, which has attribute *c,* which uses interface *C* to activate the fourth part, which has attribute *d* and which uses interface *D* to activate the fifth part, which has attribute *e,* which uses interface *E* to activate the sixth part, which has attribute *f*".

| Symbol | Meaning | Terminal |
| --- | --- | --- |
| A,B,C,... | Standard Interfaces | Non-Terminal |
| & | AND | Non-Terminal |
| ( | Begin Definition of a Sub-System (Sub-Systems can contain Sub-Systems within them.) | Non-Terminal |
| ) | End Definition of a Sub-System (Sub-Systems can contain Sub-Systems within them.) | Non-Terminal |
| A-,B-,C-,... | Negative Half of Standard Interface | Terminal |
| A+,B+,C+,... | Positive Half of Standard Interface | Terminal |
| a,b,c,... | Attributes of Parts | Terminal |

| *(whitespace)* | Separates Discrete Parts | Terminal |

So far, this code can only represent branching networks, "trees". Closed loops such as feedback loops cannot be represented. This can be solved by using numbers to represent parts in the decode order. The first part decoded as the decoder moves from left to right through the system-level code is also represented by the number 1, the second part by 2, and so forth. For example:

*aAbBcCdFe&(DfE2)*   (System-Level Code for Network of Six Parts with Feedback Loop)

represents

*aA- A+bB-E+ B+cC- C+dF-D- F+e D+fE-*  (Part-Level Code for Network of Six Parts with Feedback Loop: Part 6 activates Part 2)

In these examples, only the terminal symbols correspond directly to the coding genes. The terminal symbols probably represent functional domains within the genes and the corresponding proteins. The terminal symbols may be used in the system-level code, leading to redundant appearances of the same symbols or genes for a genetic code. The jigsaw mechanism, as explained previously, might lead to the appearance of the genetic sequences for matching proteins or matching interface domains without the start and stop sequences found in the coding genes. The rewriting that translates the system-level code, the non-terminal symbols and possibly some terminal symbols, into the part-level code, all terminal symbols, might superficially resemble random duplication and shuffling of genes and segments of genes.

4.3  SELF-REFERENTIAL SYSTEM-LEVEL CODES

The genetic code must describe the decoder for the genetic code. All living things contain instructions for the machinery used to read the genetic code and manufacture the biochemical machinery, for example the ribosome. This is similar to computer language compilers such as C language compilers that are themselves written in the computer language. The decoder is the definition of the genetic code.

A true system-level genetic code, if it exists, probably describes the decoder for the system-level genetic code in the system-level genetic code. Single-step mutations in the system-level genetic code can affect the code for the decoder and may cause large scale changes in the decoder. Ideally a mutation in the code for the decoder generates another valid decoder for a system-level genetic code. A true system-level genetic code can evolve not only the system that it describes but also the decoding machinery itself. A crude system-level genetic code, provided that it is self-referential, can evolve a more sophisticated system-level genetic code. The actual system-level code may be quite complex.

4.4  REAL-WORLD MOLECULAR BIOLOGY

The context-free-grammar-based system-level codes in this section may illustrate the basic principles of the genetic system used by eukaryotes.  The symbols in the grammars such as *A-B*, *a-b*, *A+*, and so forth may correspond to the regions delimited by the introns, the so-called exons, both in the coding genes and in the pseudogenes and other junk DNA.  The terminal symbols may correspond to the exons used in the coding genes.  The terminal symbols may also be found in the non-coding DNA such as the pseudogenes.  The non-terminal symbols correspond to exons or other regions found in the non-coding DNA that may differ substantially from the exons in the coding DNA.  The terminal symbols should correspond to functional domains in the proteins that fold independently and can be separated and recombined without changing shape or function.  The presence of the non-terminal symbols suggests that the pseudogenes should exhibit significant differences from the coding genes.  Many features matching this description have been reported in the genetic code for the blood clotting cascade and usually attributed to random gene duplication and shuffling of the exons[13].

The non-terminal symbols representing the standard interfaces, the upper case letters *A-Z* in the examples, could be arbitrary DNA sequences indexing a library of pairs of matched interfaces, the DNA sequence for a matched pair of interface components such as *(DNA sequence for interface)(marker)(DNA sequence for matched interface)*, a DNA sequence such as *(DNA sequence for a template interface component)(instructions for building a matching interface using a template mechanism)*, or *(DNA sequence for a parent protein)(DNA sequence for a jigsaw protein to cut the parent into two or more matching pieces)*.  There is little doubt that significantly more complex system-level codes than the simple examples in this section are possible.

5. SYSTEM-LEVEL GENETIC CODES, IRREDUCIBLE COMPLEXITY, AND INTELLIGENT DESIGN

The term irreducible complexity was popularized by Michael Behe in his book *Darwin's Black Box*.  The term is closely associated with the intelligent design movement which includes University of California at Berkeley law professor Phillip Johnson, Michael Behe, William Dembski and a number of others[14].  The intelligent design advocates argue that irreducible complexity implies an intelligent designer or designers.  The intelligent designer is frequently implied to be or explicitly identified with God[15].  The present paper accepts irreducible complexity as a useful concept but does not accept that irreducible complexity implies intelligence nor that intelligent design of life implies a God or gods described in any world religion.

Professor Behe's book contains at least two separate arguments that are conflated into a single position.  First, the book defines irreducible complexity and then argues that several biochemical systems including parts of the blood clotting cascade are irreducibly complex systems. Behe implicitly accepts the standard belief that each protein in his examples is coded independently by a single coding gene.  His argument that certain biochemical systems such as parts of the blood clotting cascade pose a problem for Darwinian evolution depends on this assumption.

Second, the book argues that since human beings, intelligent designers, create irreducibly complex systems including most man-made machines, therefore an intelligent designer created the irreducibly complex biochemical systems described in the book. Although this inference may be true, it is an example of a well-known logical fallacy. Given three logical propositions P (intelligent designers), Q (irreducible complexity), and R, if P implies Q is true does not imply that Q implies P because R may also imply Q. To make the case it is necessary to prove that only intelligent designers can create irreducibly complex systems. This is a logical fallacy that is frequently repeated in the intelligent design arguments.

Irreducibly complex systems appear rare in nature outside of living organisms and man-made machines. However, two and three dimensional jigsaw puzzles can be produced when a sheet or block of material is struck and shatters into matching pieces. This can happen when a rock falls through a sheet of ice on the surface of a frozen pond. Jigsaw puzzles are a toy example. Nonetheless they meet the definition of irreducible complexity. In addition, the spatial patterns formed when a solid object shatters are neither purely random nor rigidly regular as in a crystal. Complex, non-repeating patterns are formed. Other examples of irreducibly complex systems produced by natural forces may exist.

System-level genetic codes may provide a naturalistic mechanism that could plausibly evolve irreducibly complex systems through random variation and natural selection without the introduction of new physical phenomena. In addition, system-level genetic codes may provide a naturalistic mechanism for the "hopeful monsters" that have been proposed to account for major gaps in the fossil record and to leap the functional discontinuities between different types of life.

If a system-level genetic code should be identified and deciphered in living organisms, this would not explain the origin of the system-level genetic code. It would make the evolution of irreducibly complex systems through a process of random variation and natural selection substantially more plausible. The existence of a system-level genetic code would not prove that the irreducibly complex systems evolved by random variation and natural selection. The system-level genetic code might be an artifact of an intelligent designer or another process in which random variation and natural selection played little or no role – a process fundamentally different from Darwinian evolution.

The hypothetical system-level genetic code may seem miraculous if one assumes that the first living organisms used a part-level genetic code. If the first living organisms used a primitive system-level genetic code instead of a part-level genetic code, life could have then evolved more sophisticated system-level genetic codes and the part-level code that we observe today by exploiting the directed search and "jumps" provided by the system-level genetic code. This primitive system-level genetic code would probably have relied on biochemical versions of the jigsaw or template mechanisms described above.

## 6. CONCLUSIONS

System-level genetic codes may account for a variety of problems with standard evolutionary theory and probably can provide a mechanism for systemic macromutations without introducing an intelligent agent or new physical phenomena. Human design and engineering provides the best examples of system-level representations that might be used in nature. The human example suggests a template mechanism, a jigsaw mechanism, reusable standardized interfaces between parts, and also the use of master copies that are rarely if ever used in the mass production of parts. In particular, the master-copy systems widely used in design and manufacturing may explain the seemingly non-functional pseudo-genes found in many genetic systems.

The ideas in this paper strongly suggest a thorough reanalysis of the junk DNA, including the pseudogenes and the introns, especially the junk DNA associated with the coding genes for complex biochemical systems such as the blood clotting cascade that are candidates for irreducible complexity or the coding genes for biochemical systems that seem extremely complex such as metabolic pathways. The DNA associated with complex biochemical systems is most likely to contain the system-level codes if a system-level genetic code exists. Context free grammars, attribute grammars, and related concepts from computer science and linguistics are good candidates for the underlying mathematical structure that should be sought in the junk DNA.

The practical significance of deciphering the system-level genetic code if it exists may be substantial. The blood clotting cascade is a good candidate for a system in which a biochemical system-level genetic code may be present. Malfunctions of the blood clotting cascade cause some heart attacks. Blood clots have been found in the autopsies of some heart attack victims. It has been speculated that the abrupt, improper formation of blood clots in the veins and arteries feeding the heart causes the vast majority of heart attacks although the blood clots are rarely found during autopsies. Improper formation of blood clots is a common cause of death and disability. The blood clotting cascade is extremely complex and many parts remain poorly understood. Deciphering the hypothetical system-level information contain in the genetic code for the blood clotting cascade could lead to methods to prevent improper formation of blood clots such as dietary measures or new drugs. If system-level information exists, sequencing of the coding genes for the blood clotting cascade may prove unrevealing.

The early stages of the blood clotting cascade are poorly understood because of the low concentration of these proteins in the blood. This makes experiments on the early stage proteins in the cascade difficult to perform. If blood clotting causes most heart attacks (the author is skeptical of this theory), the faulty triggering of the blood clotting cascade probably occurs in this poorly understood sequence of reactions[16]. The hypothetical system-level genetic code for the blood clotting cascade would explain how this poorly understood process actually works and perhaps what chemicals in the blood, veins, or especially the plaque on the interior of the veins trigger the blood clotting process.

Cancer is a good candidate for a systemic macromutation of a single cell. In general, mutations able to produce detectable changes in organisms are negative mutations resulting in organisms that soon perish or would perish in the wild.  Most random mutations of single cells in organisms should result in cells that simply die due to a malfunction of the complex cellular machinery.  Cancer cells on the other hand become unusually robust.  Most strains that have been cultured in the laboratory become immortal and can survive without the supporting machinery of the organism[17].  Cancer cells frequently exhibit large-scale chromosomal abnormalities that suggest a large-scale restructuring of the genome[18,19].  This is difficult to reconcile with point-mutation, viral oncogene, oncogene, and anti-oncogene hypotheses. However, a mutation in the hypothetical system-level genetic code might produce these large changes in the chromosomes.

It is striking that chromosomal abnormalities exist in cancer cells at all.  Random rearrangement of man-made machines – imagine, for example, interchanging a car engine and back seat or, worse, interchanging half of a car engine and half of a back seat – invariably renders the machines non-functional and it is difficult to see how random changes to chromosomes would not produce catastrophic results.  In contrast, the point-mutation, oncogene, viral oncogene, and anti-oncogene hypotheses would be similar to the car ignition switch breaking so that the car cannot be turned off − a simple, plausible localized mutation of a single part or a few parts that could cause cancer without destroying the cell.  Yet, contrary to naive intuition, cancer cells frequently contain demonstrable chromosomal abnormalities.  This strongly suggests that the chromosomal abnormalities frequently found in cancer cells are not random changes.

The transition from single-cell to multi-cellular organisms required the creation of a control system that forced the cells to cooperate.  The system-level genetic code theory attributes this transition to a mutation in the system-level genetic code that created the control system in a single step or a small number of steps.  Cancer would be the obvious result if this control system failed, disappeared, or substantially changed due to a systemic macromutation in a single cell.  A large-scale change in chromosomal structure, whether caused by a mutation in the hypothetical system-level genetic code or not, could easily overwrite or disable the entire multi-cellular control system.  The cell would revert to an autonomous single-celled organism and devour its host.  In the system-level genetic code theory, the instructions for the multicellular control system are contained in the system-level genetic code and it is in the system-level code that the change or changes that cause cancer occur.  Cancer is a leap back across the evolutionary divide between multi-cellular organisms and the primordial single-celled organisms.

If cancer were attributed to the failure or deactivation of a regulatory gene or genes governing the multi-cellular control system then the large-scale chromosomal abnormalities frequently observed in cancer cells would not occur.  The cancer cell would differ from healthy cells by only one or a few regulatory genes.  Further, mapping of these small differences using gene sequencing technology would have rapidly identified the genetic and biochemical basis of cancer through comparison of cancer cells and healthy cells.  The regulatory genes would necessarily code for proteins that would be damaged or absent in the cancer cells.  Cancer could then be treated simply by synthesizing the protein manufactured by the undamaged genes in the healthy cells and flooding the tumor with the correct regulatory proteins.  This would restore the multi-cellular control system to healthy function inhibiting or even eliminating the tumor. This has obviously not been achieved − probably because the cancer cells differ dramatically from the healthy cells at a genetic level and the change or changes that cause cancer are difficult to identify.

If the hypothetical system-level genetic code is modified, then the entire part-level genetic code may be re-derived from the system-level description. Chromosomes could be rearranged, not in a random manner that would almost certainly produce a non-functional cell but in a highly organized manner producing a highly functional killer cell that devours its host. For example, the sections in the chromosomes rearranged in the chromosomal abnormalities are not selected at random but correspond to the genetic code for an entire sub-system such as the multi-cellular control system. The simplest explanation is that the rearrangements shut down the multi-cellular control system or a critical part of the multi-cellular control system.

If re-derivation of the part-level genetic code, the coding genes, from the system-level genetic code occurs, the re-derivation probably only occurs during cell division. Since the re-derivation can cause global systemic changes in the DNA, such as for example restructuring of chromosomes, the entire DNA must be unraveled into separate strands as occurs during cell division. The partial unraveling of small regions of the DNA double helix when the messenger RNA for isolated genes is produced will, in general, be inadequate for global changes. There is no requirement that the re-derivation of the part-level genetic code occurs during every cell division. Indeed it may occur only under specialized conditions.

In the system-level genetic code theory of cancer suggested here, the multi-cellular control system is either absent or substantially modified in cancer cells. An entire system of proteins is either absent from or greatly changed in cancer cells. This makes treating or curing cancer more difficult than a cancer due to the few defective or missing genes in conventional genetic theory. Simply replacing a defective or absent protein will not work because the proteins that react with this protein are also missing from the cancer cells. On the other hand the cancer cells will contain substantial differences from the healthy cells since an entire biochemical system is either absent or grossly modified. A "magic bullet" protein that attacks only the cancer cells should be easier to develop once the biochemical system has been identified, possibly through decoding of the system-level code for the multi-cellular control system. What is needed in this case is a destructive protein that is easily deactivated by a multi-cellular control system protein found only in the healthy cells or a proenzyme that is activated by a mutant multi-cellular control system protein found only in the cancer cells. The proenzyme is converted to a destructive enzyme that destroys the cancer cell. The proenzyme method will not work if the multi-cellular control system is absent in the cancer cells. The proenzyme might be created by crossing a digestive enzyme with a constituent of the mutant multi-cellular control system from the cancer cells.

If the cancer cell lacks one or more proteins found in healthy cells − for example, the multi-cellular control system or a significant part of the control system is missing − the "magic bullet" could be fashioned from a cascade of two or more tightly coupled proteins. With two proteins, the simplest "magic bullet", the first protein is a proenzyme for a destructive enzyme that will kill the cell such as a digestive enzyme or biological toxin. The second enzyme would be an activation enzyme for the proenzyme. The activation enzyme is engineered so that a regulatory protein found only in the healthy cells destroys the activation enzyme. Treatment would consist of first introducing the activation enzyme into the patient. The activation enzyme would be destroyed in the healthy cells but build up to a relatively high concentration in the cancer cells. Ideally the activation enzyme should affect only the proenzyme to avoid side-effects. Next, the "magic bullet" proenzyme would be introduced into the patient. The proenzyme would remain inactive in the healthy cells but would be converted to the enzyme in the cancer cells that now contain relatively high concentrations of the activation enzyme. Because the concentration of the multi-cellular control system proteins is probably quite low in both the healthy and cancer cells, the cascade may need more than two proteins to amplify the signal. The building blocks of the "magic bullet" cascade can be fashioned by cannibalizing parts of the multi-cellular control system cascade from healthy cells.

Something more sophisticated will be needed if the changes in the cancer cells change the relative frequencies of proteins or the temporal order of production of proteins in the multi-cellular control system but do not delete or modify any of the proteins or add new proteins that indirectly modify the system. For example, the changes could duplicate genes causing excess production of regulatory proteins. Alternatively, the position of the genes along the chromosome or on different chromosomes or relative to other genes or non-coding markers could control the temporal order in which proteins are synthesized. Nonetheless, these speculations illustrate how decoding the system-level genetic code if it exists could lead directly to a treatment for cancer.

Incidentally, any theory attributing cancer to a failure of the multi-cellular control system can explain the difficulty in finding the biochemical difference between cancer cells and healthy cells, the presumably missing or damaged proteins, needed for effective treatment. Control systems such as the early stages of the blood clotting cascade are information processing systems and do not require large, easily detectable concentrations of proteins to function. They act as triggers turning on or off blood clotting, cell division, and other functions. Thus it might be virtually impossible to identify the missing or damaged proteins from chemical analysis and comparison of cancer cell cultures and healthy cell cultures. If the damaged genes or the entire system of genes could be identified, the genes from healthy cells and cancer cells could be inserted into cooperative bacteria through recombinant DNA techniques and the missing or damaged proteins mass-produced for research and treatment purposes.

The aging process suggests pre-planned system-wide obsolescence. All parts in an organism seem to degrade together according to a programmed schedule that varies widely from species to species. At the extremes, some trees live thousands of years and some insects only days. If a system-level genetic code exists, the instructions governing the aging process would be found in the code. Since aging is ubiquitous among multi-cellular organisms, one must suspect that it serves some useful purpose. Caution would be needed in applying any knowledge acquired from deciphering this system.

The immune system is another candidate for a system in which a biochemical system-level genetic code may be present. Presently, a large number of serious incurable diseases are attributed to problems with the immune system. These include Acquired Immune Deficiency Syndrome (AIDS), diabetes, multiple sclerosis, rheumatoid arthritis, and lupus. The system-level genetic code should contain information about the interrelationships between the many parts of the immune system.

The biochemical system is highly interrelated and most drugs cause side-effects. The system-level genetic code would explain the interrelationships between seemingly widely separated biochemical systems. Side-effects could be predicted and hopefully drugs could be designed without side-effects or with minimal side-effects.

7. ABOUT THE AUTHOR

John F. McGowan is an engineer and researcher in the field of digital video at NASA Ames Research Center. He is Technical Lead for the Desktop Video Expert Center at NASA Ames. He has worked on digital video quality metrics, still image quality metrics, perceptual optimization of JPEG still image compression, and the conceptual design of a video system for the Mars Airplane. He has designed and implemented MPEG-1 and MPEG-2 digital audio and video software decoders. He is the author of John McGowan's AVI Overview, a popular Internet Frequently Asked Questions (FAQ) on the AVI digital audio/video file format. He worked on the Mark III and SLD experiments at the Stanford Linear Accelerator Center (SLAC). John F. McGowan has a B.S. in Physics from the California Institute of Technology and a Ph.D. in Physics from the University of Illinois at Urbana-Champaign. His personal web page is http://www.jmcgowan.com/. He can be reached at jmcgowan@mail.arc.nasa.gov or jmcgowan@veriomail.com.